\documentclass[12pt,aip,jcp]{revtex4-1}

\usepackage{graphicx}
\usepackage{amssymb}
\usepackage{amsfonts}
\usepackage{amstext}
\usepackage{amsbsy}
\usepackage{amsmath}
\usepackage{amsopn}
\usepackage{color}
\newcommand\kboltz{{k_{\!{\scriptscriptstyle \rm B}}}}
\newcommand{\nn}{\nonumber}

\begin{document}

\title{Electron-vibration energy exchange models in nitrogen-containing plasma flows}

\author{V.~Laporta\footnote{v.laporta@ucl.ac.uk}$^,$}
\affiliation{Department of Physics and Astronomy, University College London, London WC1E 6BT, UK}
\affiliation{Istituto di Metodologie Inorganiche e dei Plasmi, CNR, 70125 Bari, Italy}
\author{D.~Bruno\footnote{domenico.bruno@cnr.it}$^,$}
\affiliation{Istituto di Metodologie Inorganiche e dei Plasmi, CNR, 70125 Bari, Italy}

\begin{abstract}
The physics of vibrational kinetics in nitrogen-containing plasma produced by collisions with electrons is studied on the basis of recently derived cross sections and rate coefficients for the resonant vibrational-excitation by electron-impact. The temporal relaxation of the vibrational energy and of the vibrational distribution function is analyzed in a state-to-state approach. The electron and vibrational temperature are varied in the range of [0--50000]~K. Conclusions are drawn with respect to the derivation of reduced models and to the accuracy of a relaxation time formalism. A analytical fit of the vibrational relaxation time is given.
\end{abstract}

\maketitle

%------------------------------------------------------------------------------------------------------------------------------
\section{Introduction \label{sec:Introduction}}

Modeling of high speed flows of reactive gas mixtures is required for the description of space vehicles entries in planetary atmospheres, as well as for the interpretation of ground based experiments in high enthalpy wind tunnels and shock tubes.

It has long been recognized that vibrational energy relaxation, beside its obvious role in the global energy balance, plays a major role in the chemical kinetics of non-equilibrium zones~\cite{springerlink:10.1007/s11090-011-9339-7}. Among the many processes that influence the molecular vibrational energy, in this paper we focus on inelastic collisions with free electrons, the so-called electron-vibration (e-V) processes. Even in conditions of weak ionization, as they occur in Earth entries at orbital speed or in Mars entries, excitation by electron-impact is very efficient a mechanism for populating vibrationally excited molecules. In addition, the aerospace community shows a growing interest for flows with sensibly higher ionization degrees. This is due to the interest in superorbital speed Earth entries as they happen in the return phase from Lunar and Martian exploration missions~\cite{Surzhikov_2012, ISI:000293259600005, Shang201246}, and to the investigation of plasma flow control concepts (see Ref.~\citenum{Shang201246} for a review).

For nitrogen-containing plasma flows, the direct process of N$_2$ vibrational-excitation by electron scattering is a optically forbidden transition, due to the lack of dipole momentum for the N$_2$ molecule. For incident electron energies lower than 4~eV, however, the vibrational transition occurs \emph{via} a resonant process involving the $^2\Pi_g$ state of the N$_2^-$ ion:
\begin{equation}
e^- + \textrm{N}_2(\textrm{X}\,^1\Sigma_g^+;v) \longrightarrow \textrm{N}_2^- (^2\Pi_g) \longrightarrow e^- + \textrm{N}_2(\textrm{X}\,^1\Sigma_g^+;w)\,, \label{eq:res_scatt}
\end{equation}
where $v$ and $w$ are the initial and final vibrational level of N$_2$. By means of the resonant process in \eqref{eq:res_scatt} the vibrational-excitation cross section is enhanced by several orders of magnitude. This process is very efficient and makes the vibrational temperature equilibrate very rapidly with the free electron temperature~\cite{gorse_ricard85} to the extent that many simple models assume the N$_2$ vibrational temperature instantaneously equilibrates with the free electron temperature. A first serious theoretical study for e-V relaxation by electron-impact has been proposed by Lee~\cite{Lee_92}. More refined models have been developed by Mertens~\cite{ISI:000079859500005} and by Bourdon and Vervisch~\cite{ISI:A1997WV24900103, ISI:000090142100005}.

The experimental data on resonant vibrational-excitation cross sections and rate coefficients are difficult to obtain and, in the literature, they are limited to the first few levels~\cite{Allan}. In order to build a high temperature model information on the upper vibrational levels is mandatory. Recently, theoretical calculations have been performed for electron-nitrogen resonant scattering considering all vibrational transitions up to the dissociation limit in the energy range $0-10$~eV~\cite{0963-0252-21-5-055018} that show very good agreement with available experimental data. The new theoretical results are used in this paper to set up a kinetic study of electron-impact vibrational energy relaxation in nitrogen-containing plasma with the purpose of assessing the accuracy of existing models.

The paper is organized as follows: in section~\ref{sec:Method} the basic equations of the kinetic model are presented; in section~\ref{sec:Xsec} the full set of rate constants for all vibrational transitions are introduced and used to assess modeling assumptions commonly used in the literature. The solution of the full kinetic model is analyzed in section~\ref{sec:Solution_Meq} together with a reduced multi-quantum vibrational transition approach. The relaxation time formalism is introduced in section~\ref{sec:Relaxation} and the results of its application to the vibrational energy relaxation are compared to those obtained within the full kinetic model. Section~\ref{sec:Conclusions} summarizes the main conclusions and perspectives.

%------------------------------------------------------------------------------------------------------------------------------
\section{Modeling of electron-vibration relaxation \label{sec:Method}}

In order to model the e-V kinetics a nitrogen gas in contact with a free electron bath is considered. The electrons are considered at equilibrium at the constant temperature $T_e$ and the N$_2$ molecules are supposed to be in the electronic ground state $\textrm{X}\,^1\Sigma^+_g$, with a initial vibrational distribution at temperature $T_v$. Due to the very efficient resonant mechanism, the electron-N$_2$ collisions drive the vibrational distribution towards the equilibrium at the electron temperature $T_e$. Since the electron affinity of the N atom is negative, the resonant dissociative electron attachment for electron-N$_2$ is suppressed~\cite{0963-0252-21-5-055018}. Therefore, the effects of molecular dissociation on vibrational relaxation are not taken into account.

The time evolution of the molecular vibrational distribution function (VDF) is described by the following equations:
\begin{equation}
\frac{\partial n_v}{\partial t} = n_e \sum_{w}{\left[ K_{w,v}\, n_w - K_{v,w}\, n_v \right]}\,,
\label{eq:vibr_kin_eqs}
\end{equation}
where $n_e$ is the electron density and $K_{v,w}(T_e)$ is the rate constant for the $v \rightarrow w$ resonant vibrational-excitation (RVE) process:
\begin{equation}
K_{v,w}(T_e) = \frac{2}{\sqrt\pi}\,(\kboltz T_e)^{-3/2}\,\int\,\epsilon\,\sigma_{v,w}(\epsilon)\,e^{-\epsilon/\kboltz T_e}\,d\epsilon\,, \label{eq:rate_coeff}
\end{equation}
where $\sigma_{v,w}$ is the corresponding cross section as a function of the electron energy $\epsilon$. In Eq.~\eqref{eq:vibr_kin_eqs} the sum extends over all 67 vibrational levels of N$_2$. The detailed balance principle imposes the following condition among the direct and inverse processes:
\begin{equation}
K_{v,w}\,n_v^* = K_{w,v}\,n_w^*\,, \label{eq:detailed_bil}
\end{equation}
where $n_{v}^*$ is the equilibrium population of the vibrational level $v$ at temperature $T_e$:
\begin{equation}
n_v^* \sim e^{-\epsilon_v/\kboltz T_e}\,, \label{eq:nveq}
\end{equation}
$\epsilon_v$ being the energy of the N$_2$ vibrational level $v$ and $\kboltz$ the Boltzmann constant.
The total N$_2$ vibrational energy is readily obtained from the VDF:
\begin{equation}
E_{vib} = \sum_{v}{n_v\, \epsilon_v}\,.\label{eq:Evib}
\end{equation}

%------------------------------------------------------------------------------------------------------------------------------
\section{Rate coefficients \label{sec:Xsec}}

In this section the rate coefficients $K_{v,w}(T_e)$ that enter the kinetic equations~(\ref{eq:vibr_kin_eqs}) and (\ref{eq:Evib}) for the resonant process in Eq.~(\ref{eq:res_scatt}) are presented. They are taken from Ref.~\citenum{0963-0252-21-5-055018} (paper I in the following) where full details on the calculations and the validation of the model by comparison with other approaches can be found. The method used in I to calculate the electron-N$_2$ resonant scattering is the well-known `boomerang model', an approximation of the nonlocal-complex-potential model~\cite{wadehra}. Such a method has been used to study many different systems and a complete description together with the application limits can be found in the papers~\cite{0963-0252-21-5-055018, wadehra, 0963-0252-21-4-045005} and references therein. Molecular rotations have been taken into account by adding the centrifugal barrier to the potential, parameterized by the rotational quantum number $J$ and transitions with $\Delta J=0$ have been considered. Results show that the rate constants for vibrational-excitation do not depend strongly on rotational excitation so, in the following, only $J=0$ is considered.

\begin{figure}[t!]
\begin{center}
\includegraphics[scale=.7,angle=0]{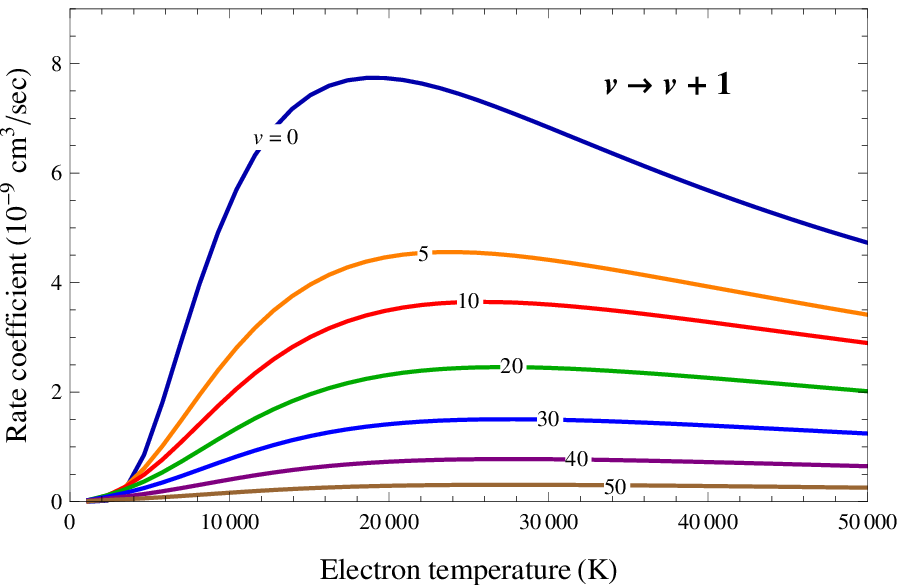}
\includegraphics[scale=.7,angle=0]{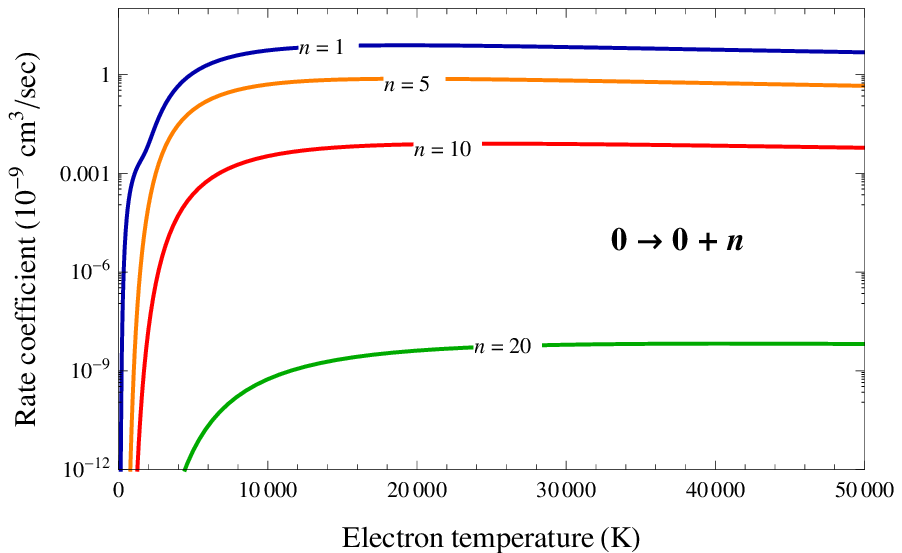}
\caption{Theoretical rate coefficients for electron-N$_2$ resonant vibrational excitations  as a function of electron temperature as obtained in Ref.~\citenum{0963-0252-21-5-055018}. Left: Mono-quantum transitions; Right: $n$-quantum transitions starting from the ground vibrational level. \label{fig:rateN2}}
\end{center}
\end{figure}

In I, a full set of RVE cross sections among all vibrational levels of N$_2$ has been calculated. The corresponding rate coefficients have been obtained by averaging each cross section over a equilibrium electron energy distribution as defined in Eq.~\eqref{eq:rate_coeff} in the range [0--50000]~K of electron temperature. Figure~\ref{fig:rateN2} shows typical rate constants for mono-quantum transitions ($v\to v+1$) and for multi-quantum transitions starting from the ground vibrational level ($0\to0+n$).

\begin{figure}[t]
\begin{center}
\includegraphics[scale=.7,angle=0]{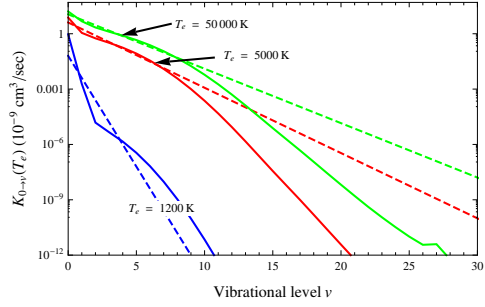}
\caption{Rate coefficients $K_{0,v}$ as a function of the final vibrational level $v$ calculated at different electron temperatures. Dashed lines are linear extrapolations obtained using the first 5 points. \label{fig:Mertens}}
\end{center}
\end{figure}

Since a large number of transitions are involved in the full data set, it is natural to look for regularities that allow using scaling relations. On the other hand, in the past both theoretical and experimental studies on RVE by electron-impact were, as a rule, limited to a small subset of all possible processes: that required the introduction of scaling-laws among the rate coefficients in order to obtain the missing data. Mertens~\cite{ISI:000079859500005} proposed to assume a linear relation between $\log ( K_{0,v} )$ and the final vibrational level $v$. Bourdon \emph{et al.}~\cite{ISI:000090142100005} have shown, using the Allan data~\cite{Allan}, that this rule is approximately valid up to $v=7$ at $T_e=1$~eV. In Fig.~\ref{fig:Mertens} we report the calculated rate coefficients $K_{0,v}$ as a function of the final vibrational level $v$ for different electron temperatures $T_e$, together with a linear extrapolation obtained using the first 5 points. A linear relation holds with good accuracy for levels $v \leqslant 5$. At higher electron temperatures this limit can be raised to $v \approx 10$ but transitions with larger jumps are overestimated by many orders of magnitude. As we shall see in Sec.~\ref{sec:Solution_Meq}, transitions with quantum jumps up to $\Delta v_{max} = 20$ have to be considered for a accurate estimation of the vibrational energy relaxation.

\begin{figure}[t]
\begin{center}
\includegraphics[scale=.7,angle=0]{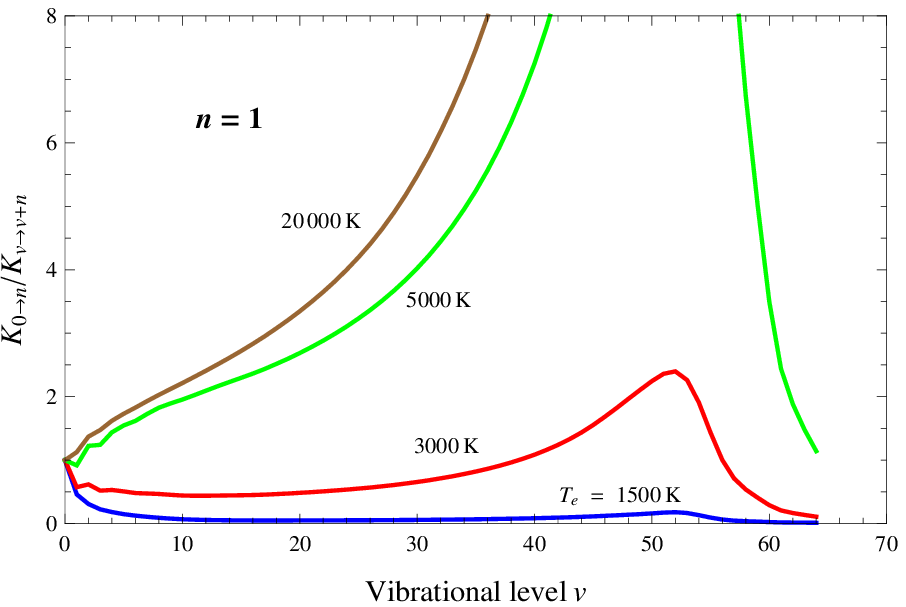}
\includegraphics[scale=.7,angle=0]{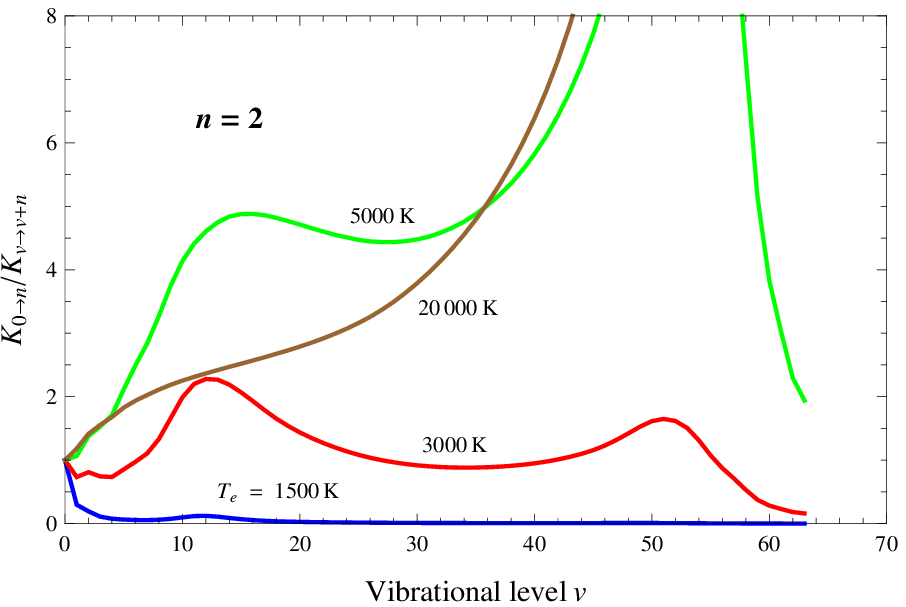}
\caption{$K_{0,n}/K_{v,v+n}$ as a function of the vibrational level $v$, for different values of the electron temperature $T_e$, for mono-quantum (left) and bi-quantum (right) transitions. \label{fig:Gordiets}}
\end{center}
\end{figure}

For transitions starting from vibrational levels other than the ground state, Gordiets~\cite{ISI:A1995RW96300029} proposed the following scaling law:
\begin{equation}
K_{v,v+n} = \frac{K_{0,n}}{1+ a\,v}\,, \label{eq:Gordiets}
\end{equation}
$a$ being an adjustable parameter. In Fig.~\ref{fig:Gordiets} the ratio $K_{0,n}/K_{v,v+n}$ as a function of the initial vibrational level $v$, for different values of the electron temperature and for different values of the vibrational jump $n$, is reported. The results show that the linear relation, Eq.~\eqref{eq:Gordiets}, holds approximately only in limited regions of the parameters' space and that the $a$ parameter is temperature-dependent.

The scaling-laws proposed by Gordiets and Mertens are heuristic guesses coming from gas discharge physics~\cite{gorse_ricard85} where typical conditions are electron temperature around 1~eV and the gas temperature below 1000~K. Figures~\ref{fig:Mertens} and \ref{fig:Gordiets} show that indeed linear scaling-laws are reasonable assumptions when the electron temperature is in the range where the rate coefficients have a broad maximum ($10000 \lesssim T_e \lesssim 30000~\textrm{K}$ see Fig.~\ref{fig:rateN2}) and for vibrational levels up to $v=10$. For high speed flows applications, however, models should be valid in a much broader range of the parameters' space. The linearity deteriorates for low electron temperatures and high vibrational levels and extrapolation of these relations can lead to severe overestimation of the rate constants. In the following, only the rate coefficients from paper I are used.

%------------------------------------------------------------------------------------------------------------------------------
\section{Vibrational energy relaxation \label{sec:Solution_Meq}}

\begin{figure}[t!]
\begin{center}
\includegraphics[scale=.7,angle=0]{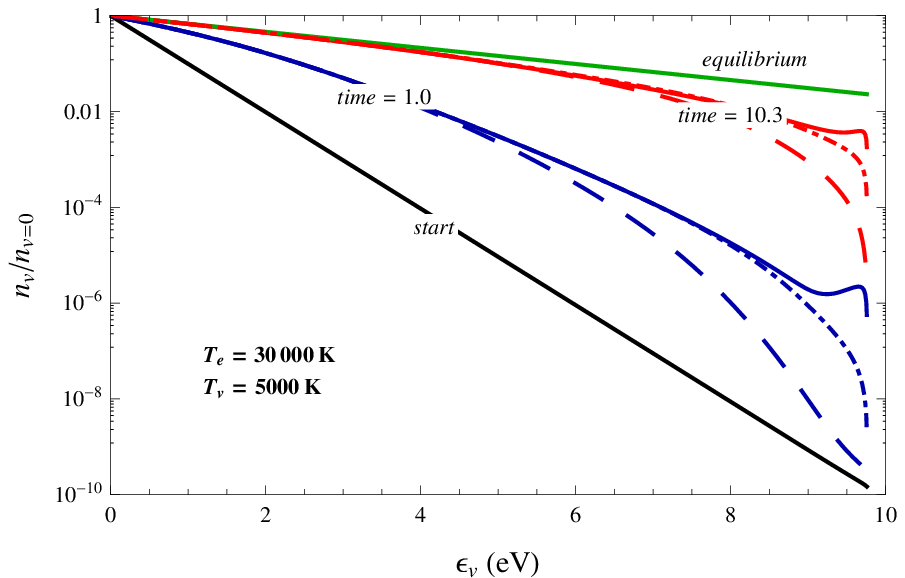}
\includegraphics[scale=.7,angle=0]{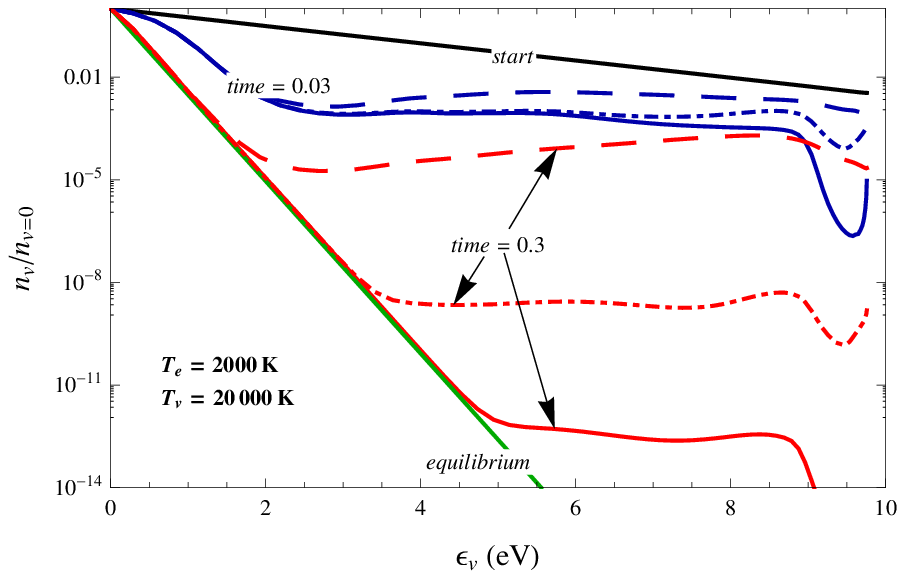}
\caption{Temporal evolution of normalized vibrational distributions function as obtained by different assumptions on the maximum number of vibrational quanta exchanged in collisions. Full line: all transitions allowed; dashed line: $\Delta v_{max}=10$; dot-dashed line: $\Delta v_{max}=20$. Left panel: $T_e=30000$~K; $T_v=5000$~K. Right panel: $T_e=2000$~K; $T_v=20000$~K. \label{fig:MJvdf}}
\end{center}
\end{figure}

\begin{figure}[t!]
\begin{center}
\includegraphics[scale=.7,angle=0]{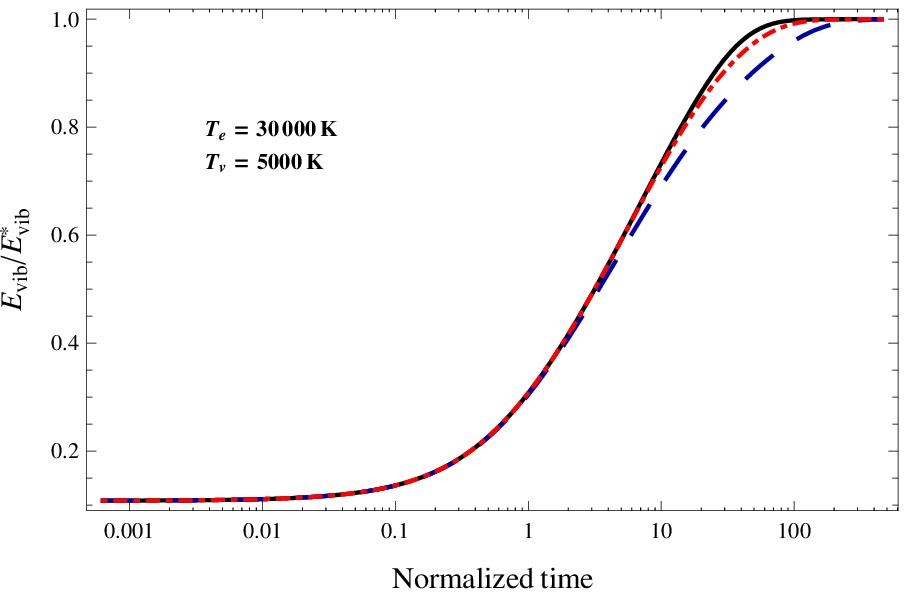}
\includegraphics[scale=.7,angle=0]{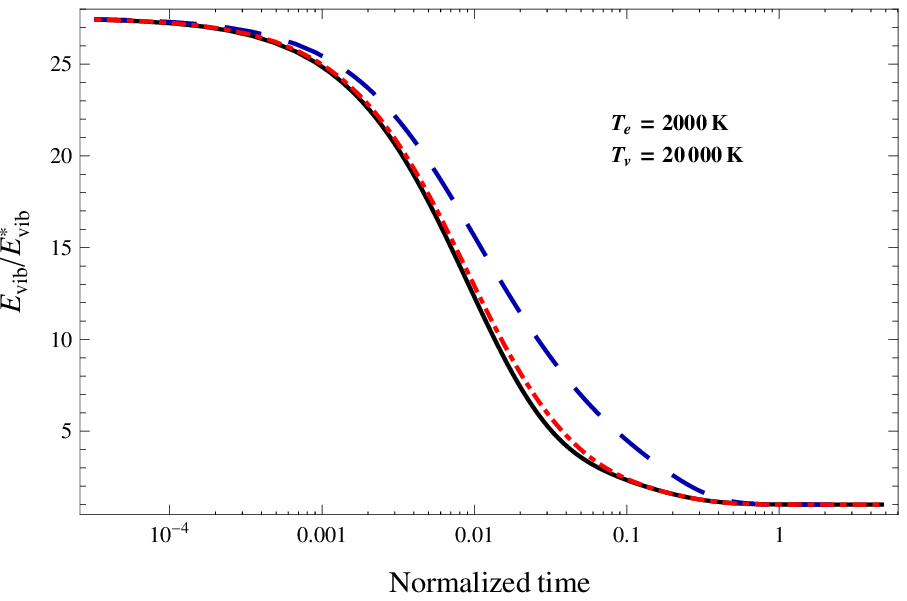}
\caption{Temporal evolution of the normalized vibrational energy obtained by different assumptions on the maximum number of vibrational quanta exchanged in collisions. Full line: all transitions included; dashed line: $\Delta v_{max} = 10$ levels; dot-dashed line: $\Delta v_{max} = 20$ levels. Left panel: $T_e=30000$~K; $T_v=5000$~K. Right panel: $T_e=2000$~K; $T_v=20000$~K. \label{fig:MJev}}
\end{center}
\end{figure}

Using the rate coefficients presented in section~\ref{sec:Xsec}, the time evolution of the VDF of N$_2$ molecules and of the vibrational energy $E_{vib}$ is studied by solving the coupled rate equations in Eqs.~\eqref{eq:vibr_kin_eqs}. In particular the role of the multi-quantum transitions is investigated. The initial condition for the VDF is supposed to be a Boltzmann distribution at temperature $T_v$. At given electron temperature, the rate constants for transitions involving the exchange of many vibrational quanta are much smaller than the others due to the large energy gap: see for example Fig.~\ref{fig:rateN2}. It is therefore natural to ask wether it is possible to set a upper limit to the number of quanta exchanged in order to reduce the amount of data required for the modeling. Figures~\ref{fig:MJvdf} and \ref{fig:MJev} show the `full' solution, \textit{i.e.} obtained including the full set of multi-quantum transitions, and the effect of considering a restricted range of vibrational transitions in the vibrational kinetics on the VDF and on the corresponding $E_{vib}$ for two typical conditions of heating ($T_v=5000$~K and $T_e=30000$~K) and cooling ($T_v=20000$~K and $T_e=2000$~K). VDF is normalized to the population of the vibrational ground state and vibrational energy $E_{vib}$ is normalized to its equilibrium value $E^*_{vib}$. Times are normalized by:
\begin{equation}
\tau_{01} (T_e)= \frac{1}{n_e K_{0,1}(T_e)}\,, \label{eq:tau01}
\end{equation}
so that the plots are independent of electron and molecular number densities. Figure~\ref{fig:MJvdf} shows the relaxation of the VDF. The latter, although a Boltzmann distribution at the beginning (by construction) and at the end (by the equilibrium condition) of the relaxation, shows non-equilibrium character both in heating and cooling conditions. In both cases, high vibrational levels relax more slowly than low ones and the VDF exhibits strong depletion (overpopulation) of high levels with respect to a Boltzmann distribution at the same total energy. Large multi-quantum jumps have a strong influence on this relaxation and restricting the maximum number of exchanged vibrational quanta overestimates the degree of non-equilibrium by many orders of magnitude. The plot on the right in Fig.~\ref{fig:MJvdf} shows that this error can involve vibrational levels as low as $v \sim 10$. It is interesting to note that a fairly good agreement is found in Fig.~\ref{fig:MJev} for vibrational energy, even if significant discrepancies are observed on the VDFs, as vibrational energy is mostly due to low vibrational levels.

\begin{figure}[t!]
\begin{center}
\includegraphics[scale=.7,angle=0]{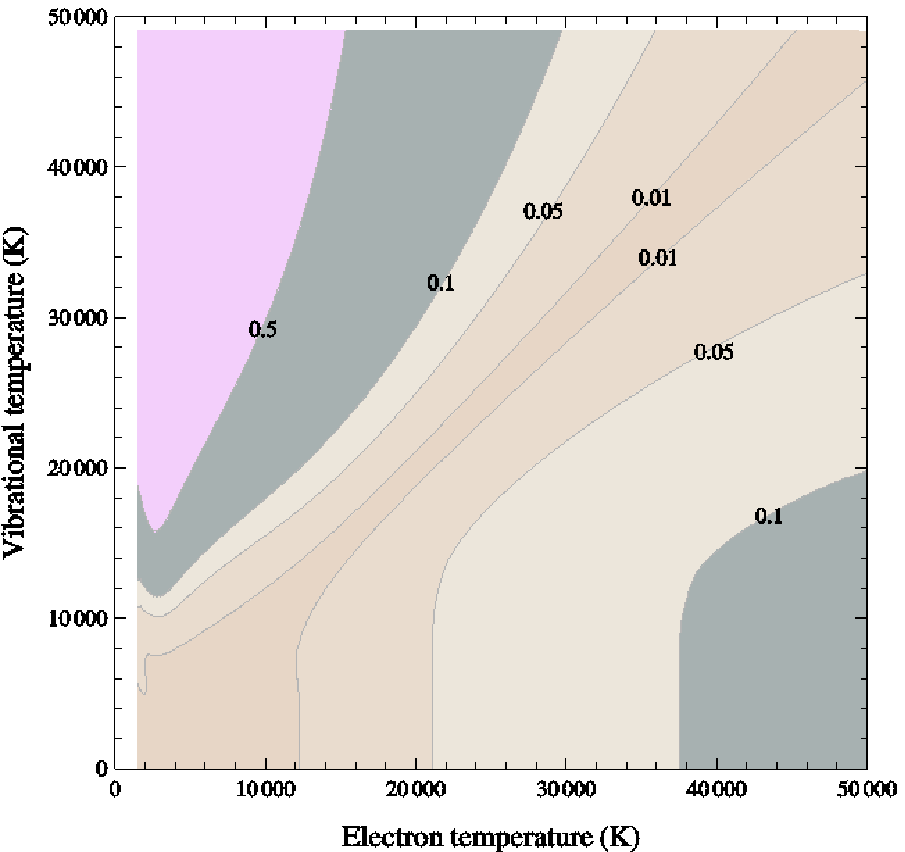}
\includegraphics[scale=.7,angle=0]{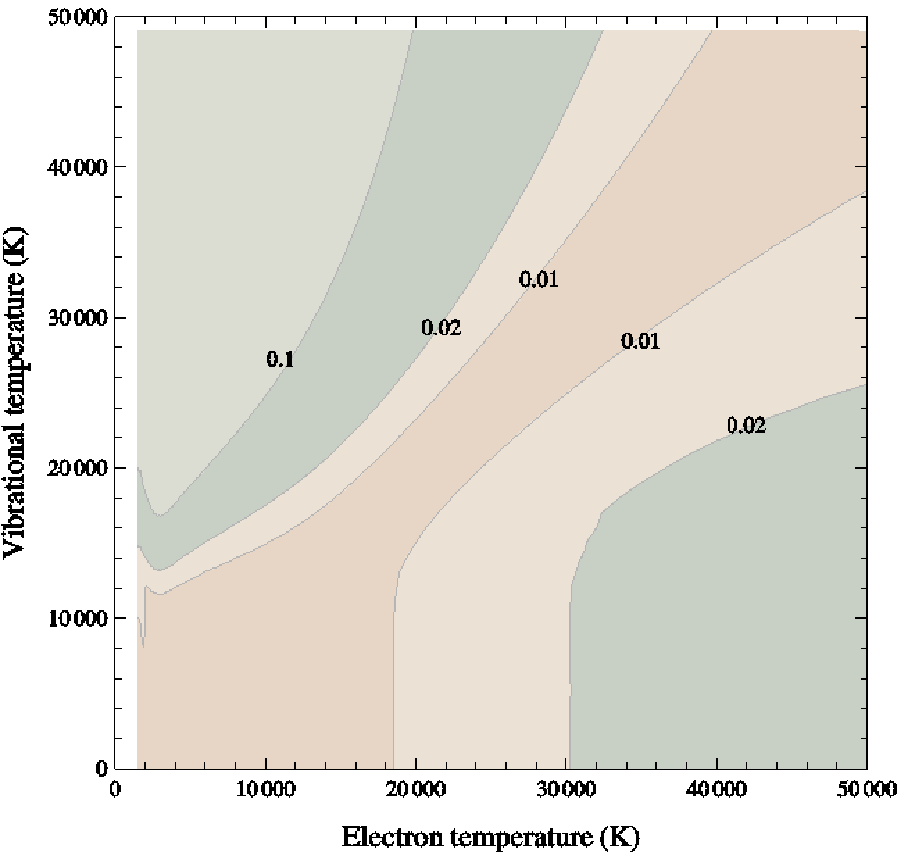}
\caption{Maximum error during the temporal evolution of $E_{vib}$ obtained by considering multi-quanta transitions with $\Delta v_{max} = 10$ (left) and $\Delta v_{max} = 20$ (right). \label{fig:MultiJumps}}
\end{center}
\end{figure}

Figure~\ref{fig:MultiJumps} reports the maximum error, calculated over all the temporal evolution of $E_{vib}$, between the results obtained by considering the full set of rate coefficients and those obtained including only transitions with $\Delta v_{max} = 10$ and $\Delta v_{max} = 20$ in the kinetic model. Results show $\Delta v_{max} = 10$ is a reasonable approximation with a maximum error smaller than 10$\%$ in a wide region of the parameters' space. Large errors appear for cooling conditions ($T_v > T_e$) when the starting vibrational temperature is large enough. Although so large vibrational temperatures are difficult to realize in practice, this is an indication that very large multi-quantum jumps are involved in the relaxation  of strongly populated high vibrational levels. With $\Delta v_{max}=20$ the maximum error stays below 1\% for most conditions of interest.

The vibrational energy relaxation is therefore described with reasonable accuracy by taking into account only transitions where less than 10 vibrational quanta are exchanged; if a accurate description of the vibrational distribution is required, however, at least $\Delta v_{max} = 20$ must be considered.

\begin{figure}[t!]
\begin{center}
\includegraphics[scale=.7,angle=0]{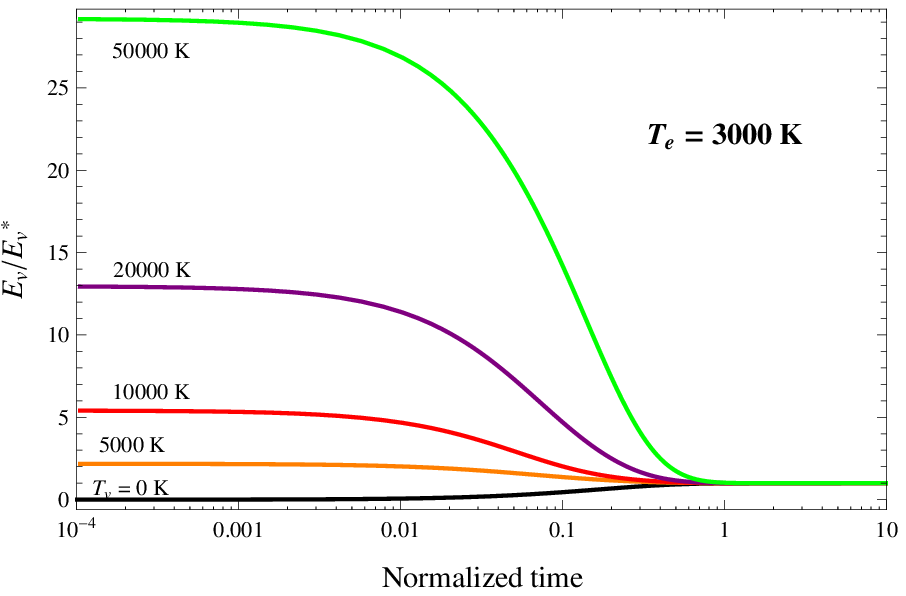}
\includegraphics[scale=.7,angle=0]{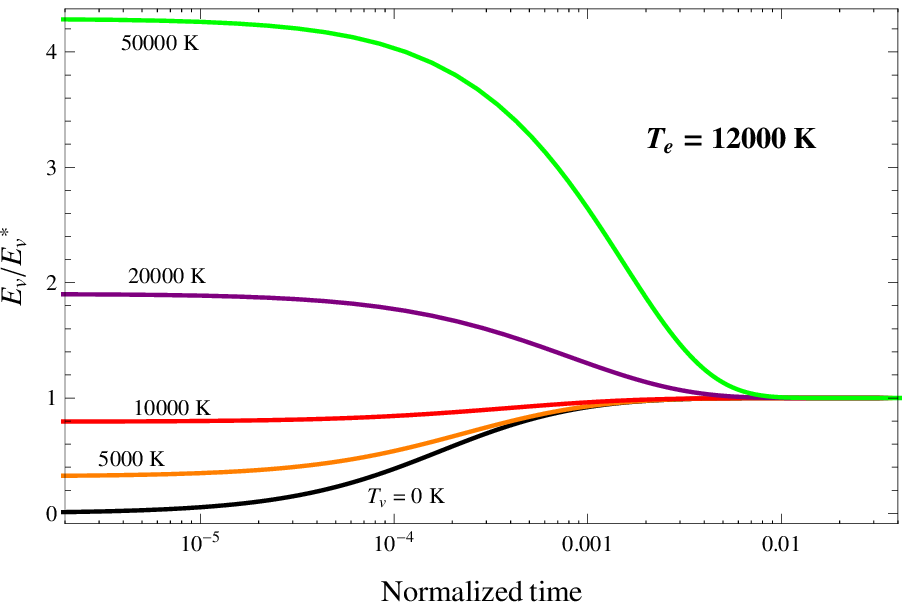}
\caption{Time evolution of the normalized vibrational energy, $E_{vib}/E^*_{vib}$, for two different values of the electron temperature $T_e$, parameterized on the N$_2$ initial vibrational temperature $T_v$. \label{fig:ver_tetv}}
\end{center}
\end{figure}

To conclude this section, Fig.~\ref{fig:ver_tetv} summarizes the temporal evolution of the vibrational energy, as obtained by the full kinetic model, for two electron temperatures 3000~K and 12000~K and for initial vibrational temperature in the range $T_v=0$ to 50000~K to model different heating and cooling conditions. It is clear that the rate of relaxation is mainly determined by the electron temperature. But the initial vibrational distribution also plays a role since different initial conditions relax along different paths.

%------------------------------------------------------------------------------------------------------------------------------
\section{Relaxation time formalism \label{sec:Relaxation}}

The standard way of modeling the vibrational energy relaxation is to describe it by a Landau-Teller (LT) rate equation:
\begin{equation}
\frac{d E_{vib}}{dt} = \frac{E^*_{vib} - E_{vib}}{\tau_e}\,, \label{eq:Landau-Teller}
\end{equation}
where $E^*_{vib}$ is the equilibrium vibrational energy at the electron temperature $T_e$, and $\tau_e$ is the e-V relaxation time. Early theoretical work of Lee~\cite{Lee_92} studied the relaxation from a $T_v=0$ condition assuming a harmonic oscillator model for the N$_2$ molecule and using the cross sections from Ref.~\citenum{Huo}. Results can be summarized with the following analytical expression for $\tau_e$:
\begin{equation}
p_e \tau_e^{\textrm{Lee}}= \kboltz T_e \left\{ \frac12 \left(\frac{\theta_v}{T_e} \right)^2  \sum_v K_{0,v}(T_e)\, v^2 \right\}^{-1}\,, \label{eq:tauLee}
\end{equation}
where $p_e=n_e\kboltz T_e$ is the electron partial pressure and $\theta_v$ is the characteristic vibrational temperature of the N$_2$ molecule. The relaxation time defined in Eq.~\eqref{eq:tauLee} depends only on $T_e$ and it does not take into account vibrational excitation. Lee's formulation has been tested and improved in the works of Mertens~\cite{ISI:000079859500005} and Bourdon \emph{et al.}~\cite{ISI:A1997WV24900103, ISI:000090142100005}. In these works, cross sections have been taken from the theoretical work by Huo \emph{et al.}~\cite{Huo} or from the experimental data of Allan~\cite{Allan}. The Huo \emph{et al.} results are \textit{ab-initio} calculations for $0 \rightarrow v \leqslant 5$ transitions with rotational quantum numbers $J=0, 50, 150$. The corresponding rate coefficients are calculated in the electron temperature range $0.1-5.0$~eV. In addition, results have been obtained for $J=50$ and initial vibrational quantum number up to $v=12$ with changes in vibrational quantum number of $-5$ and $+5$. In particular, the results obtained for $J=50$ have been used in the modeling calculations~\cite{ISI:000079859500005, ISI:000090142100005}. The Allan's results are experimental values for $0 \rightarrow v$ transitions up to $v=13$ obtained at room temperature. In both cases, cross sections for all other transitions must be obtained by assuming a scaling-law~\cite{ISI:000079859500005, ISI:A1995RW96300029}. As a result, analytical expressions have been derived for the relaxation time in Eq.~\eqref{eq:Landau-Teller} that depends both on $T_e$  and $T_v$~\cite{ISI:000079859500005, ISI:A1997WV24900103, ISI:000090142100005}.

Analyzing the full set of RVE rate coefficients, we concluded in Sec.~\ref{sec:Xsec} that such scaling-laws have strong limitations. It is therefore interesting to investigate the limits of a LT description of the vibrational energy relaxation and to provide, where possible, updated values for the relaxation time.
To this end, time histories of vibrational energy for different conditions of electron temperature $T_e$ and initial vibrational temperature $T_v$ have been calculated with the full kinetic model. As it will shown in the following, in general, the relaxation is not a linear process. Therefore some arbitrariness is implicit in the definition of the relaxation time. In this paper the value of $\tau_e$ was obtained in each case by linearizing the relaxation rate around its starting point, in order to minimize the discrepancies between the LT and the full model values for the vibrational energy.

\begin{figure}[t!]
\begin{center}
\includegraphics[scale=.7,angle=0]{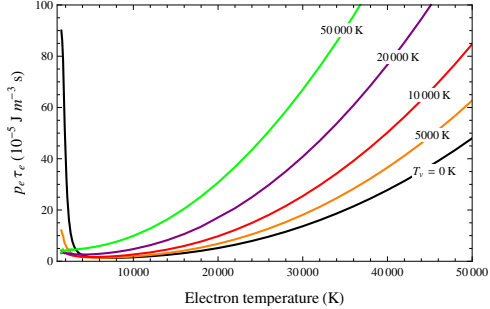}
\caption{Vibrational relaxation time as a function of electron temperature for different values of the initial vibrational temperature $T_v$. The values are obtained by solving the full set of kinetics equations. \label{fig:relaxtime}}
\end{center}
\end{figure}

\begin{figure}[t!]
\begin{center}
\includegraphics[scale=.7,angle=0]{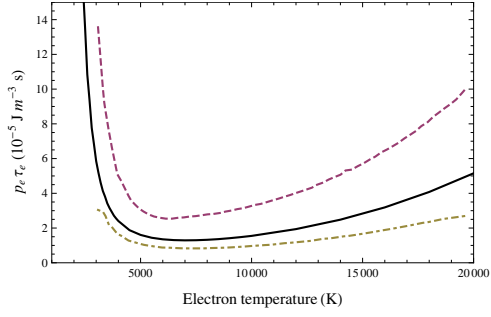}
\caption{Vibrational relaxation time as a function of the electron temperature with initial vibrational temperature $T_v=0$~K. Full line: this work; dashed line: results from Ref.~\citenum{Lee_92}; dotted line: results from Ref.~\citenum{ISI:000090142100005}.\label{fig:relaxtime_cfr}}
\end{center}
\end{figure}

The final results are reported in Fig.~\ref{fig:relaxtime} and in particular in Fig.~\ref{fig:relaxtime_cfr} they are compared with those of Lee~\cite{Lee_92} and Bourdon \emph{et al.}~\cite{ISI:000090142100005} for $T_v=0$. Present results lie between the other two, somewhat closer to the more recent results of Bourdon \emph{et al.}

\begin{figure}[t!]
\begin{center}
\includegraphics[scale=.7,angle=0]{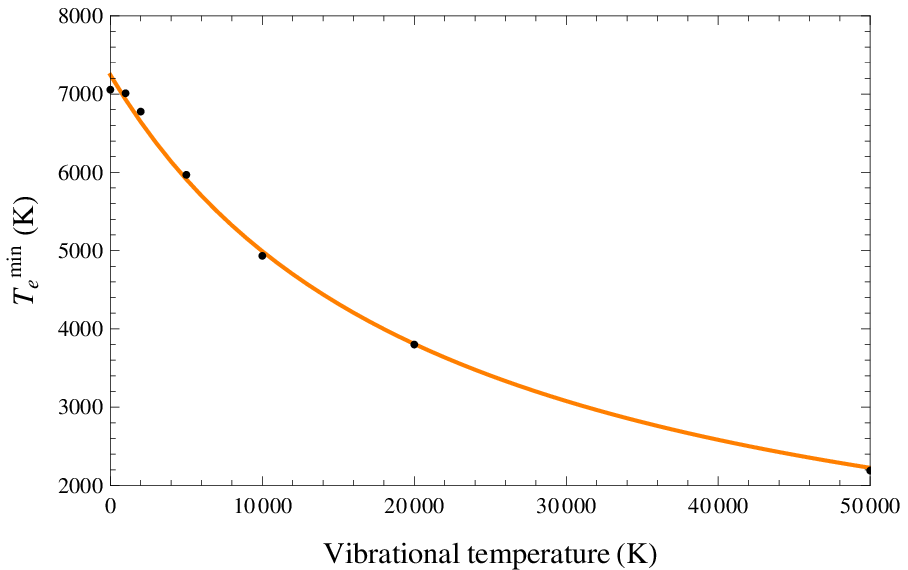}
\includegraphics[scale=.7,angle=0]{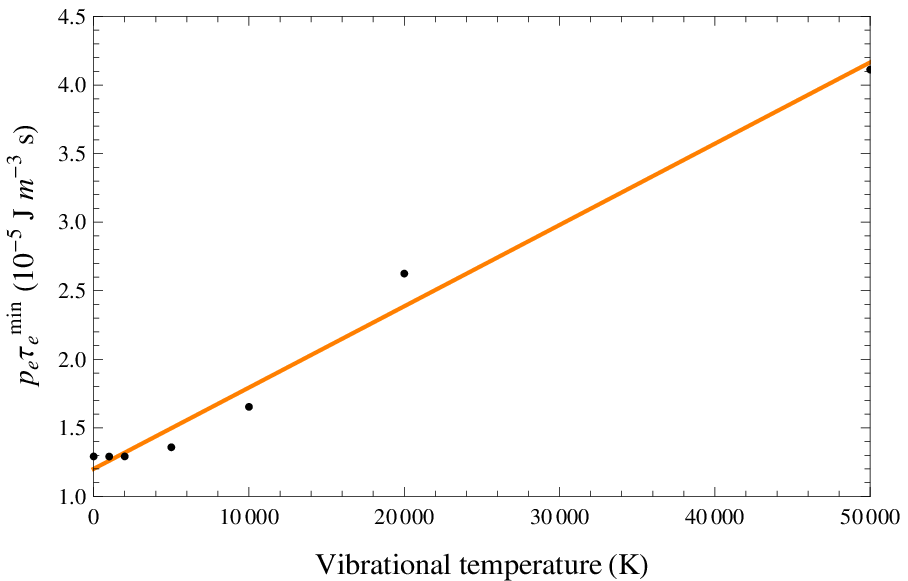}
\caption{Best fit of $T_e^{\textrm{min}}$ and $p_e\tau_e^{\textrm{min}}$ as a function of the initial vibrational temperature, as given in Table~\ref{tab:Te_min}. \label{fig:Te_min}}
\end{center}
\end{figure}

Due to the resonant character of the cross sections, the relaxation time shows a minimum. In Table~\ref{tab:Te_min} the point of the minimum $(T_e^{\textrm{min}},\, p_e\tau_e^{\textrm{min}})$ of the relaxation time, for some vibrational temperatures, is shown. These points can be fitted by the following laws:
\begin{eqnarray}
T_e^{\textrm{min}}(T_v) &=& \frac{b}{T_v+a}\,,\label{eq:fit_Temin}
\\
p_e\tau_e^{\textrm{min}}(T_v) &=& c + d\,T_v\,,\label{eq:fit_petaue}
\end{eqnarray}
where the coefficients take the following values:
\begin{eqnarray}
a &=& 22159.90~\textrm{K}\,,\nn
\\
b &=& 1.60\times 10^8~\textrm{K}^2\,,\nn
\\
c &=& 1.20\times 10^{-5}~\textrm{Jm}^{-3}\textrm{s}\,,\nn
\\
d &=& 5.9\times10^{-10}~\textrm{Jm}^{-3}\textrm{sK}^{-1}\,. \label{eq:fit_Temin_coeff}
\end{eqnarray}
Figure~\ref{fig:Te_min} shows the best fit superposed to the calculated points.
\begin{table}[t!]
\begin{center}
\begin{tabular}{ccc}
\hline\hline
~~$T_v$ (K)~~ & ~~$T_e^{\textrm{min}}$ (K)~~ & ~~$p_e\tau_e^{\textrm{min}}$ (10$^{-5}$ Jm$^{-3}$s)~~ \\
\hline
0     & 7055.64 & 1.29 \\
1000  & 7010.24 & 1.29 \\
2000  & 6776.13 & 1.29 \\
5000  & 5968.45 & 1.35 \\
10000 & 4932.81 & 1.65 \\
20000 & 3800.00 & 2.62 \\
50000 & 2189.37 & 4.11 \\
\hline\hline
\end{tabular}
\caption{Minimum of relaxation time $(T_e^{\textrm{min}},\, p_e\tau_e^{\textrm{min}})$, for some initial vibrational temperatures $T_v$. \label{tab:Te_min}}
\end{center}
\end{table}

For electron temperature $T_e\geqslant T_e^{\textrm{min}}$ the relaxation time curves in Fig.~\ref{fig:relaxtime} can be fitted, with $1\%$ accuracy, by the following polynomial expression:
\begin{equation}
p_e\tau_e(T_e,T_v) = p_e\tau_e^{\textrm{min}}(T_v) + \sum_{i=1}^3 c_i(T_v)\,\left(T_e-T_e^{\textrm{min}}(T_v)\right)^i\,, \hspace{2cm} T_e\geqslant T_e^{\textrm{min}} \label{eq:petaue_fit}
\end{equation}
where the coefficients $c_i(T_v)$ are given by:
\begin{eqnarray}
c_1(T_v) &=& 2.38\times10^{-5} - 4.54\times10^{-9}\,T_v + 1.65\times10^{-13}\,T_v^2\,,\nn
\\
c_2(T_v) &=& 1.78\times10^{-8} + 2.11\times10^{-12}\,T_v - 2.23\times10^{-17}\,T_v^2\,,\nn
\\
c_3(T_v) &=& 1.17\times10^{-13}\,.\label{eq:fit_taue_coeff}
\end{eqnarray}

\begin{figure}[t!]
\begin{center}
\includegraphics[scale=.7,angle=0]{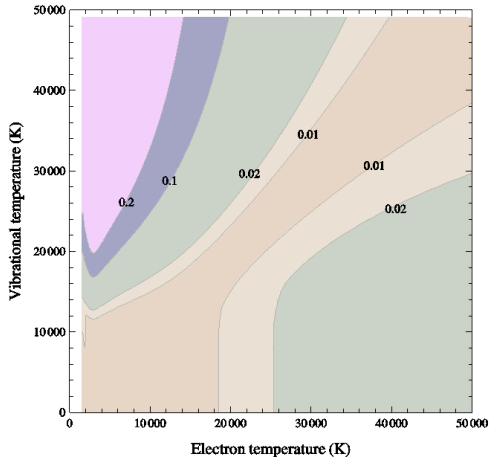}
\caption{Maximum deviation on the vibrational energy as obtained with a Landau-Teller approach compared with that obtained by solving the full vibrational kinetic model.\label{fig:maxerr}}
\end{center}
\end{figure}

The LT equation, Eq.~\eqref{eq:Landau-Teller}, can be integrated analytically to give the vibrational energy relaxation. For each set of initial conditions $(T_e,T_v)$, a value for the relaxation time is calculated and kept constant during all relaxation; the resulting vibrational energy and the corresponding value obtained by solving the full set of kinetic equations were compared and the maximum deviation was estimated. Figure~\ref{fig:maxerr} show the results of this analysis as a contour map. A single relaxation time, as in LT approach, is able to describe the e-V process both in relaxing and exciting conditions when the non-equilibrium is not very strong ($T_e \approx T_v$). The range of validity increases for larger $T_e$. Under strong non-equilibrium conditions, however, large discrepancies (larger than 10\%) are seen.

\begin{figure}[t!]
\begin{center}
\includegraphics[scale=.7,angle=0]{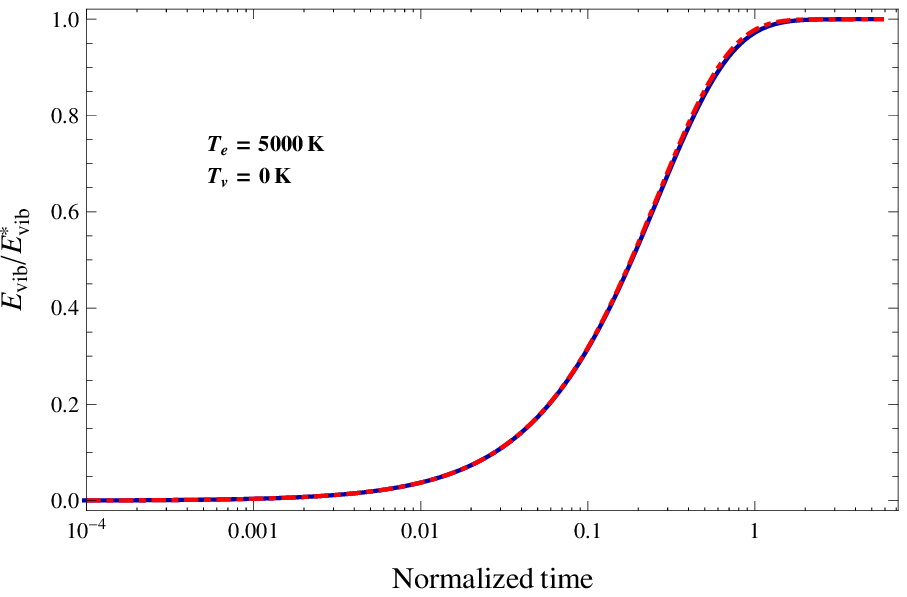}
\includegraphics[scale=.7,angle=0]{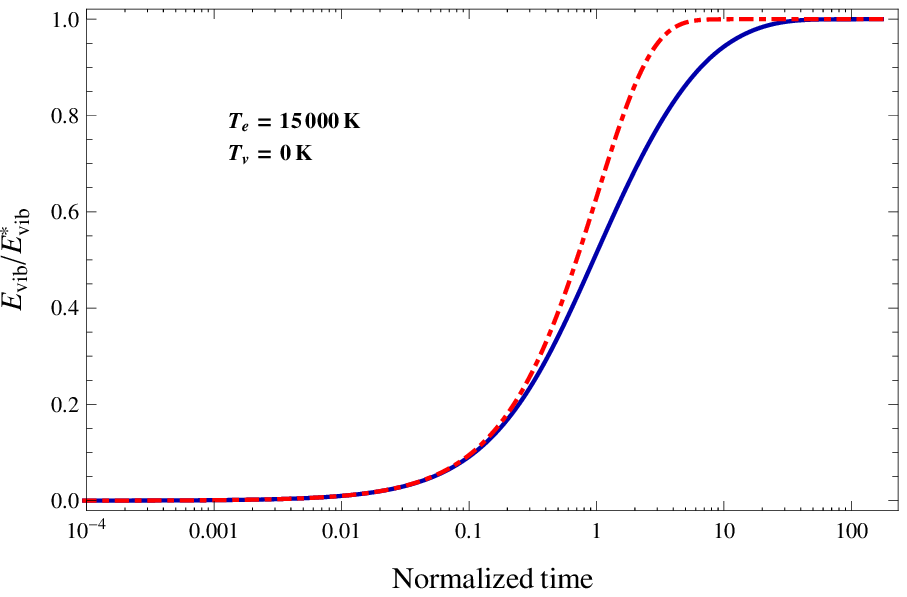}
\includegraphics[scale=.7,angle=0]{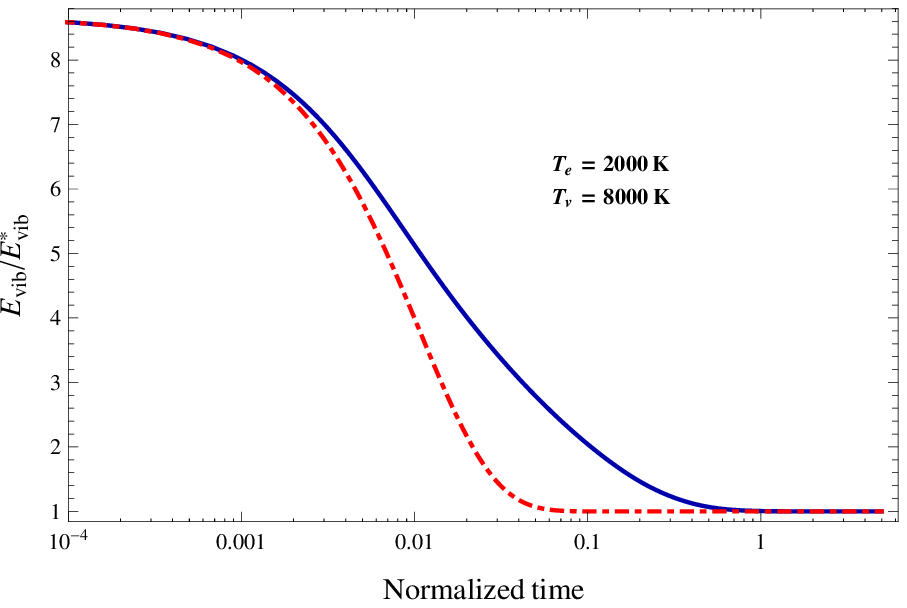}
\caption{Time relaxation of normalized vibrational energy for the cases described in Table~\ref{tab:ver_cfr}. Full line: exact solution as obtained by solving the full set of kinetic equations; dashed line: constant relaxation time in a Landau-Teller approach. \label{fig:ver_Evib}}
\end{center}
\end{figure}

\begin{figure}[t!]
\begin{center}
\includegraphics[scale=.7,angle=0]{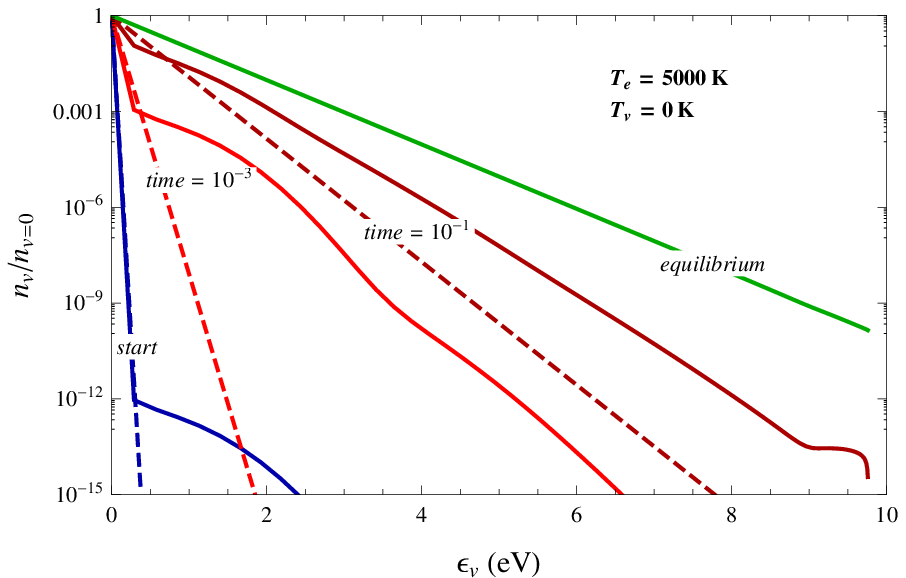}
\includegraphics[scale=.7,angle=0]{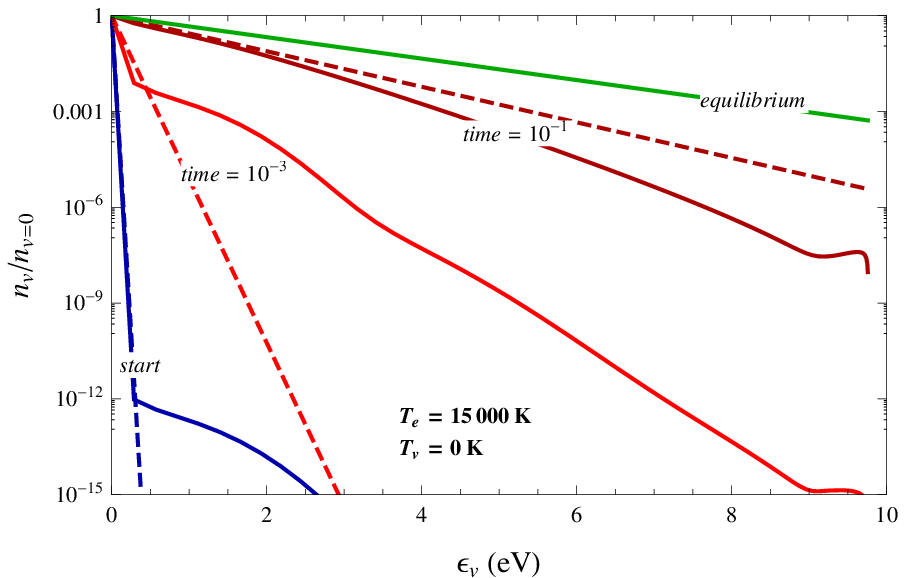}
\includegraphics[scale=.7,angle=0]{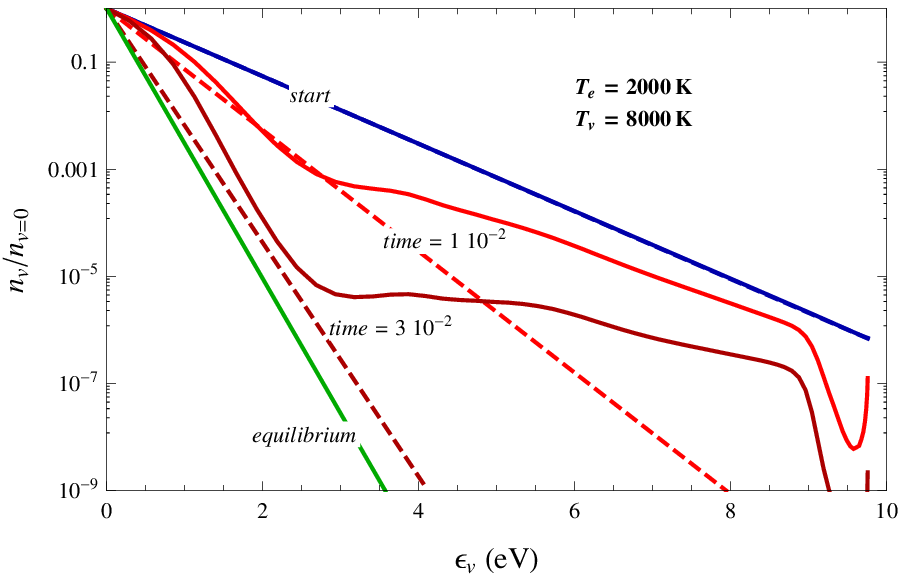}
\caption{Relaxation of the N$_2$ vibrational distribution function, normalized to the ground vibrational state, for the cases described in Table~\ref{tab:ver_cfr}. Full line: exact solution as obtained by solving the full set of kinetic equations; dashed line: constant relaxation time in a Landau-Teller approach. \label{fig:ver_vdf}}
\end{center}
\end{figure}

In order to explain the results of Fig.~\ref{fig:maxerr}, the relaxation process was analyzed in greater detail for three cases. Table~\ref{tab:ver_cfr} summarizes the conditions for the cases studied, the value for the relaxation time and the computed error with respect to the exact solution. Figure~\ref{fig:ver_Evib} shows the time evolution of the normalized vibrational energy for two heating conditions with $T_v=0$~K and $T_e=5000$ and 15000~K and a cooling condition with $T_v=8000$~K and $T_e=2000$~K. We note that the relaxation predicted by the LT equation is faster than the exact solution both in exciting and relaxing conditions. Figure~\ref{fig:ver_vdf} shows the corresponding normalized VDF at different times during the relaxation. For the LT model, Boltzmann distributions at all times during the relaxation are assumed with vibrational temperature calculated from vibrational energy. In exciting conditions, at the beginning, high vibrational levels ($v > 1$) relax faster than low levels and get strongly overpopulated. In this conditions, the LT predictions agree well with the `full' result since the low-lying levels make the largest contribution to $E_{vib}$. At later stages, the reverse is true: again the LT results agree with the `full' results for the first two levels, but now all the rest of the VDF is overestimated. This is what makes the LT value for $E_{vib}$ larger. Note that also for case 1, where the relaxation predicted by the LT equation agrees well (within $2\%$) with the `full' result, the VDF are strongly non-equilibrium during most of the relaxation. This can have a strong influence when a vibrationally favored chemical reaction (\emph{e.g.} dissociation) is considered. In relaxing conditions, instead, the high-energy levels relax slower than low levels. As a result, the LT relaxation is faster than the `full' value. Again, strongly non-equilibrium VDF appear during most of the relaxation.
\begin{table}[t!]
\begin{center}
\begin{tabular}{cccccc}
\hline\hline
~~case~~ & ~~$T_e$ (K)~~ & ~~$T_v$ (K)~~ & ~~$\tau_e$ (s)~~  & ~~$\epsilon_{max}$~~  \\
\hline
$1$ & $5000$ & $0$   & $0.023$  & $0.01$  \\
$2$ & $15000$   & $0$ & $0.013$ & $0.26$  \\
$3$ & $2000$ & $8000$ & $0.14$ & $0.62$  \\
\hline\hline
\end{tabular}
\caption{Definition of the cases treated in the text. $\tau_e$ is calculated by linearizing the energy relaxation rate obtained from the full model and $\epsilon_{max}$ is the error on $E_{vib}$ . \label{tab:ver_cfr}}
\end{center}
\end{table}

This is the origin of the failure of a relaxation time description. Obviously, the modeling done in Refs.~\citenum{Lee_92} has been improved in Refs.~\citenum{ISI:000079859500005,ISI:000090142100005} by deriving a relaxation time that depends both on $T_e$ and $T_v$.

Based on the results presented in this work for the relaxation of the VDF, however, we do not expect the inclusions of the dependence of the relaxation time on $T_v$ to significantly improve the predicting capabilities of the LT model. Indeed, in Fig.~\ref{fig:ver_tetv} we show the $E_{vib}$ relaxation for two values of $T_e$ and for several values of $T_v$. It is apparent that these quantities alone are not able to describe the system relaxation towards equilibrium.

%------------------------------------------------------------------------------------------------------------------------------
\section{Concluding remarks \label{sec:Conclusions}}

Recent calculations of RVE cross sections in nitrogen covering all vibrational transitions~\cite{0963-0252-21-5-055018} up to the dissociation limit have been used to study the vibrational energy relaxation as produced by these resonant processes.  Analysis of the full set of rate constants shows that approximate scaling-laws adopted in the past have a limited range of validity.

Also, the possibility to model the relaxation by a linear rate equation with a single relaxation time has been investigated. The results from a relaxation time formalism and from the full kinetic model are compared to conclude that the former may be used with reasonable accuracy ($\sim$10\%) to model the vibrational energy relaxation in most cases of practical interest. A analytical fit of the calculated relaxation time, that depends both on free electron and vibrational temperatures, has been provided in Eqs.~(\ref{eq:fit_Temin})--(\ref{eq:fit_taue_coeff}) that could be of interest for the modeling of this important relaxation process.

State-to-state kinetic modeling, however, predicts strongly non-equilibrium vibrational distributions during the relaxation both in heating and cooling conditions.  Although not considered here, this can have a strong influence on the chemical kinetics of reactive plasma flows, in particular where vibrationally favored reactions (\emph{e.g.} dissociation) play a role.

The next step will be the inclusion, in the relaxation kinetic of molecular rotation and of the dissociation reaction.

Finally, it is worth mentioning that the original cross section data allow the modeling of RVE processes also in conditions where the electron energy distribution function deviates significantly from equilibrium. For the present study only rate coefficients have been used. These are obtained from the cross sections by averaging with a Maxwell distribution for the electron velocities.

%------------------------------------------------------------------------------------------------------------------------------
\section*{Acknowledgments}
The authors wish to thank Prof. M. Capitelli (Universit\`{a} di Bari, Italy) for careful reading of the manuscript and helpful discussions. The research leading to these results has received funding from the European Community's Seventh Framework Programme (FP7/2007-2013) under grant agreement n\ensuremath{^\circ} 242311.

%------------------------------------------------------------------------------------------------------------------------------
%Bibliography
%\bibliographystyle{is-unsrt}{}
%\bibliography{refs}

%------------------------------------------------------------------------------------------------------------------------------

\end{document}